\documentclass[11pt,a4paper]{article}
\usepackage{jheppub} 
\usepackage{nameref}
\usepackage{dsfont}
\usepackage{mathrsfs, amssymb, amsmath, amsfonts, latexsym, graphicx}  
\usepackage[utf8]{inputenc}
\usepackage{soul}
\usepackage{physics}
\usepackage{accents}
\usepackage[T1]{fontenc}
\usepackage{makecell}
\usepackage{tabularx}
\usepackage{makecell}
\usepackage{array}

\newcommand{\expvals}[1]{\langle #1  \rangle}

\makeatletter
\DeclareRobustCommand{\loplus}{\mathbin{\mathpalette\dog@lsemi{+}}}

\usepackage{float}
\usepackage{mathtools}
\newcommand{\defeq}{\vcentcolon=}

\newcommand{\dog@lsemi}[2]{\dog@semi{#1}{#2}{270,90}}
\newcommand{\dog@semi}[3]{%
  \begingroup
  \sbox\z@{$\m@th#1#2$}%
  \setlength{\unitlength}{\dimexpr\ht\z@+\dp\z@\relax}%
  \makebox[\wd\z@]{\raisebox{-\dp\z@}{%
    \begin{picture}(1,1)
    \linethickness{\variable@rule{#1}}
    \roundcap
    \put(0.5,0.5){\makebox(0,0){\raisebox{\dp\z@}{$\m@th#1#2$}}}
    \put(0.5,0.5){\arc[#3]{0.5}}
    \end{picture}%
  }}%
  \endgroup
}
\newcommand{\variable@rule}[1]{%
  \fontdimen8  
  \ifx#1\displaystyle\textfont3\else
    \ifx#1\textstyle\textfont3\else
      \ifx#1\scriptstyle\scriptfont3\else
        \scriptscriptfont3\relax
  \fi\fi\fi
}
\makeatother
\usepackage{tikz-cd}

\usetikzlibrary{arrows.meta,calc}

\usepackage{cleveref}
\usepackage{pict2e}
\usepackage{diagbox}
\usepackage{hyperref}

\newcommand{\beq}{\begin{eqnarray}}
\newcommand{\eeq}{\end{eqnarray}}
\newcommand{\beqn}{\begin{eqnarray}}
\newcommand{\eeqn}{\end{eqnarray}}

\newcommand{\RR}{\mathbb{R}}

\usepackage{pict2e}

\newcommand{\chkM}{{\color{red} \,\checkmark\kern-5pt{}_{M}}}

\newcommand{\ee}{\end{equation}}
\newcommand{\bea}{\begin{eqnarray}}
\newcommand{\eea}{\end{eqnarray}}

\usepackage{BOONDOX-cal}

\usepackage{scalerel}
\usepackage{stackengine,wasysym}

\newenvironment{Align}{\begin{equation}
\begin{aligned}}
{\end{aligned}
\end{equation}}
\newenvironment{Align*}{\begin{equation*}
\begin{aligned}}
{\end{aligned}
\end{equation*}\par}

\usepackage{color}

\usepackage{layout}

\newcommand{\hX}{\hat{X}}
\newcommand{\hD}{\hat{D}}

\DeclareFontFamily{OT1}{rsfs}{}
\DeclareFontShape{OT1}{rsfs}{m}{n}{ <-7> rsfs5 <7-10> rsfs7 <10->rsfs10}{} 
\DeclareMathAlphabet{\mycal}{OT1}{rsfs}{m}{n}

\DeclareFontFamily{OMS}{cmsy}{}
\DeclareFontShape{OMS}{cmsy}{m}{n}{ <5><6><7><8><9> gen * cmsy <10><10.95><12><14.4><17.28><20.74><24.88> cmsy10 }{}
\DeclareMathAlphabet{\mycaltwo}{OMS}{cmsy}{m}{n}
\DeclareFontFamily{U}{MnSymbolC}{}
\DeclareSymbolFont{MnSyC}{U}{MnSymbolC}{m}{n}
\DeclareFontShape{U}{MnSymbolC}{m}{n}{
    <-6>  MnSymbolC5
   <6-7>  MnSymbolC6
   <7-8>  MnSymbolC7
   <8-9>  MnSymbolC8
   <9-10> MnSymbolC9
  <10-12> MnSymbolC10
  <12->   MnSymbolC12}{}
\DeclareMathSymbol{\intprod}{\mathbin}{MnSyC}{'270}
\AtBeginDocument{%
  \setlength{\parindent}{1.5em}%
  \setlength{\parskip}{6pt plus 1pt minus 1pt}%
}
\title{Quantum Geometry from Area Fluctuations}
\author[a,b]{Jerzy Kowalski-Glikman}
\author[a]{, Ludovic Varrin}

\affiliation[a]{National Centre for Nuclear Research, Pasteura 7, 02-093 Warsaw, Poland}
\affiliation[b]{Faculty of Physics and Astronomy, University of Wroclaw, Pl. Maksa Borna 9, 50-204
Wroclaw}
\emailAdd{jerzy.kowalski-glikman@ncbj.gov.pl}
\emailAdd{ludovic.varrin@ncbj.gov.pl}

\begin{document}
\abstract{We construct a quantum-statistical analogue of Einstein's fluctuation argument for black-body radiation in the context of causal-diamond geometry. Starting from the phase space of a stretched horizon inside a Minkowski causal diamond, we quantize the Poisson algebra generated by the fields averaged over stretched-horizon time. We then compute the fluctuations of the averaged area density of the transverse two-spheres in a thermal state constructed in analogue with the black-body thermal state. In the null limit, where the stretched horizon approaches the causal-diamond boundary, this yields a thermal fluctuation formula of the boundary area operator that contains two terms, in direct analogue with the black-body radiation. The term quadratic in the expectation value is the ``classical'' contribution, while the linear term has the Verlinde--Zurek scaling characteristic of independent microscopic constituents. In direct analogue with Einstein's interpretation of black-body energy fluctuations as evidence for light quanta, we interpret the linear area-fluctuation term as a statistical signature of discrete quanta of geometry. This provides bottom-up evidence for quantum area degrees of freedom and supports the embadon picture of null quantum geometry.}
\maketitle

\newpage
\section{Introduction}

A recurring expectation in quantum gravity is that the smooth geometry of general relativity is an effective, coarse-grained description of more elementary microscopic degrees of freedom. This idea appears in different forms across several approaches. In loop quantum gravity, area and volume operators have discrete spectra, and quantum geometry is carried by spin-network states \cite{Rovelli:1994ge,Rovelli:1995ac,Ashtekar:1996eg,Ashtekar:1997fb,Seiberg:1999vs}. In simplicial approaches, such as Regge calculus and causal dynamical triangulations, spacetime geometry is described in terms of elementary building blocks whose continuum limit is expected to reproduce general relativity \cite{Regge:1961px,Ambjorn:2004qm,Ambjorn:2010cdt}. Another notable approach is causal set theory, where the microscopic structure of spacetime is postulated to be a locally finite partially ordered set \cite{Bombelli:1987aa}. In string theory, the extended nature of strings, high-energy scattering, dualities, and spacetime uncertainty relations suggest that arbitrarily sharp localization in a classical continuum geometry is not operationally meaningful \cite{Gross:1987kza,Amati:1988tn,Yoneya:2000bt}. More recently, null and corner approaches to quantum gravity have emphasized localized boundary degrees of freedom, including area-carrying null constituents, as natural microscopic candidates for quantum geometry \cite{Freidel:2015gpa,Ciambelli:2024swv}. While these are concrete examples of quantum gravity theories that realize a notion of discrete geometry (either as a consequence of the framework or by construction), the notion of a minimal length goes back to the thought experiment of Mead \cite{Mead:1964zz} (see also \cite{Hossenfelder:2012jw} for a review), who realized that the Heisenberg uncertainty principle obstructs distance measurements below the Planck scale. This is a general argument that should be valid at least in some sense in any theory of quantum gravity.
This brings us to the following question: Is there a general argument for the existence of quanta of geometry that does not require a complete description of the ultimate theory that describes quantum gravity?\par
Statistical physics and thermodynamics provides the natural framework to extract information about a partially unknown underlying system.
A paradigmatic example is Einstein's fluctuation argument for black-body radiation. In the first paper of his \textit{annus mirabilis}, Einstein introduced the light-quantum hypothesis through an entropy-fluctuation argument for black-body radiation in the Wien regime \cite{Einstein:1905tem} (see also \cite{DuncanJanssen}, \cite{duncan2007} for an in-depth historical discussion). He showed that the probability for monochromatic radiation of energy $E$ to be localized in a subvolume $V$ has the same volume dependence as the probability for $n=E/h \nu$ independent particles to be found in that subvolume. This suggested that high-frequency radiation behaves thermodynamically as if it consisted of $n$ independent quanta of energy $h\nu$. A sharper fluctuation-theoretic formulation was given later, in 1909, when Einstein showed that the mean-square energy fluctuation of black-body radiation contains a particle-like term, proportional to the mean energy, and a wave-like term, proportional to the square of the mean energy \cite{einstein1909gegenwartigen}. In the Wien limit the particle term dominates, whereas in the Rayleigh–Jeans limit the wave term dominates. This is a remarkable line of reasoning as it provides insights into the quantum nature of light without having access to the full underlying quantum theory.\par
While quantum electrodynamics is now one of physics's most well understood theories, we are now facing a similar situation to the early 20th century physicists when it comes to quantum gravity. Recently, there have been efforts into tackling the issue from a bottom-up perspective, trying to extract insights into a quantum theory of gravity without the necessity of giving a complete top-down theory. These inquiries have led to the \textit{corner proposal} \cite{Freidel:2015gpa,Freidel:2020xyx,Freidel:2020svx,Ciambelli:2021vnn,Freidel:2021cjp,Ciambelli:2021nmv,Ciambelli:2022cfr,Ciambelli:2022vot,Freidel:2023bnj,Ciambelli:2023bmn,Ciambelli:2024swv,Ciambelli:2024qgi,Varrin:2024sxe,Neri:2025fsh,Varrin:2025okc,Kowalski-Glikman:2025hom,Ciambelli:2025ztm,Varrin:2026bmb}. Einstein's argument about the quantum nature of light fits perfectly into this philosophy. The aim of this work is therefore to develop an analogueous argument for quantum geometry. We ask whether horizon area fluctuations can contain a statistical signature of microscopic geometric constituents, in the same way that black-body energy fluctuations contain a signature of photons. \par 
This fluctuation viewpoint is also directly connected to the Verlinde--Zurek program on observable spacetime fluctuations \cite{Verlinde:2019ade,Zurek:2022snowmass}. There, a variance of the form
\begin{equation}\label{eq:vzscaling}
\expvals{\Delta \hat{A}^2} \propto \ell_{\rm QG}^2 \langle \hat A\rangle
\end{equation}
is interpreted as a quantum-gravitational area fluctuation of causal-diamond boundaries \cite{Bub:2024qmc,Ciambelli:2025flo}. Although controlled by a microscopic area scale, such fluctuations need not remain confined to microscopic distances. Their association with null boundaries can generate long-distance correlations in the geometry, leading to an effective UV/IR mixing in interferometric observables \cite{Zurek:2020vac,Li:2022geo}. This is the mechanism by which Planckian, or more generally quantum gravity, area fluctuations may become accessible to precision interferometers. Theoretical realizations of this idea have been developed using AdS/CFT modular fluctuations \cite{Verlinde:2019xfb}, near-horizon conformal descriptions \cite{Banks:2021jwj}, and shockwave geometries \cite{Verlinde:2022hhs}. Its phenomenology has been studied in the context of next-generation gravitational-wave detectors \cite{Bub:2023qgb}, image-blurring constraints \cite{Lee:2023aib}, and proposed dedicated experiments such as GQuEST \cite{Vermeulen:2024gquest}. A complementary spectral formulation of geometric noise in interferometers was
recently developed in \cite{Freidel:2026hed}.
\par
An important, but often overlooked, feature of the Verlinde--Zurek relation is that the scaling of area fluctuations is state dependent. Equation~\eqref{eq:vzscaling} should not be understood as an operator-level statement, but rather as a property of the chosen family of states. This is directly analogueous to the black-body case. As we will review below, the Wien, or particle-like, scaling of the energy fluctuations is realized in \textit{coherent states}, whose photon-number statistics are Poissonian. In contrast, it is absent in Fock states with fixed occupation number. The thermal state combines these two behaviours, containing both the linear particle-like contribution and the quadratic wave-like contribution to the fluctuation. In this context, the purpose of the present work is to construct a quantum gravitational thermal state that similarly contains both the Verlinde--Zurek scaling and the quadratic term associated with a wave-like behaviour of the gravitational field.
\par
To make this question precise, we consider the gravitational phase space associated with a stretched horizon inside a causal diamond in Minkowski spacetime. Before turning to the details of the construction, we summarize the main results.
Following \cite{Ciambelli:2025flo}, we use the covariant phase space to analyse the gravitational system consisting of a stretched horizon inside a causal diamond in Minkowski spacetime. There, the area density of the horizon's spherical cross section, together with its normal derivative, form an affine algebra when averaged over the stretched horizon time. In the null limit where the stretched horizon hugs the boundary of the causal diamond, the area density field corresponds to the length of the causal diamond squared. This finite dimensional Poisson subalgebra is associated with the coadjoint orbits of the affine group. The quantum theory is then defined on the positive unitary irreducible representation of that group. Using so-called intelligent states, which saturate the Robertson--Schrödinger uncertainty relation, we construct a thermal state in close analogy with the state describing black-body radiation in quantum electrodynamics. We then compute the fluctuations of the area density operator which, in the null boundary limit, gives the following squared fluctuations of causal diamond boundary area operator
\begin{equation}\label{eq:Afluct}
    \expvals{\Delta \hat{A}^2}_\beta = 2\pi\gamma \expvals{\hat A}_\beta + \expvals{\hat A}_\beta^2,
\end{equation}
where $\gamma$ is a constant parameter, carrying dimensions of area, that labels different families of intelligent states. The above formula is in direct analogue with the fluctuations of the energy in the black-body radiation
\begin{equation}
    \expvals{\Delta \hat{H}_{\mathrm{rad}}^2}_\beta = \hbar\omega \expvals{\hat{H}_\mathrm{rad}}_\beta + \expvals{\hat{H}_{\mathrm{rad}}}_\beta^2,
\end{equation}
where $\hat{H}_{\mathrm{rad}}$ is the radiative Hamiltonian of a single mode of electromagnetic radiation of frequency $\omega$. Comparing the two formulas, the parameter $\gamma$ can be given an interpretation of being the fundamental quantum of area, similarly to how $\hbar\omega$ defines the quantum of energy. The quadratic term in \eqref{eq:Afluct} is the expected fluctuation behaviour of waves, analogueous to the Rayleigh--Jeans regime of the black-body radiation. It dominates the fluctuations when the classical area is large when compared to the microscopic fluctuation scale $\gamma$. 
In contrast, the first term in \eqref{eq:Afluct} is a Verlinde--Zurek linear scaling of the squared fluctuation as a function of the expectation value. It dominates the fluctuations when the classical average area is small compared to the microscopic fluctuation of area associated with $\gamma$, i.e. in the deep UV regime. It is the analogue of the Wien regime and is the characteristic signature of Brownian-like statistics of independent quanta of area. The existence of this term in the fluctuations of the area operator therefore presents a strong argument that spacetime is quantized at deep UV scales.
This result is the analogue of Einstein's light quanta argument. We give the dictionary between the black-body analysis and the gravitational case in Table \ref{table:analogue}.  

\begin{table}[H]
\centering
\renewcommand{\arraystretch}{1.4}

\resizebox{\textwidth}{!}{
\begin{tabular}{ |c|c|c| } 
\hline
& Black-body & Quantum gravity \\
\hline

\makecell{Classical\\observable}
& Radiative energy: $H_{\mathrm{rad}}$
& Area density: $a$ \\
\hline
\makecell{Quantum\\observable}
& Heisenberg Hamiltonian: $\hat{H}_{\mathrm{rad}}$
& Affine position operator: $\hat{X}$ \\
\hline

\makecell{Quantum state\\reproducing\\classical quantity \\ and Wien-VZ scaling}
& \makecell{
Coherent states: $\ket{\alpha}$ \\[0.4em]
$\mel{\alpha}{\hat{H}_{\mathrm{rad}}}{\alpha} = H_{\mathrm{rad}}$,
\, $\expvals{\Delta \hat{H}_{\mathrm{rad}}^2}_\alpha \propto H_{\mathrm{rad}}$}
& \makecell{
Intelligent states: $\ket{\gamma;a,b}$\\[0.4em]
$\mel{\gamma;a,b}{\hX}{\gamma;a,b} = a$,
\, $\expvals{\Delta \hat{X}^2}_{\gamma,a,b}\propto a$
} \\
\hline

\makecell{Thermal state's\\probability density}
& $\rule{0pt}{4.5ex}
P_\beta(H_{\mathrm{rad}}) 
= \frac{1}{\expvals{H_{\mathrm{rad}}}_\beta}
e^{-\frac{H_{\mathrm{rad}}}{\expvals{\hat{H}_{\mathrm{rad}}}_\beta}}$
& $\rule{0pt}{4.5ex}
P_\beta(a) 
= \frac{1}{\expvals{\hat X}_\beta}
e^{-\frac{a}{\expvals{\hat{X}}_\beta}}$ \\
\hline

\makecell{Fluctuation\\formula}
& $\expvals{\Delta \hat{H}_{\mathrm{rad}}^2}_\beta 
= \hbar \omega \expvals{\hat{H}_\mathrm{rad}}_\beta 
+ \expvals{\hat H_{\mathrm{rad}}}_\beta^2$
& $\expvals{\Delta \hat{X}^2}_\beta 
= \frac{\gamma}{2} \expvals{\hat X}_\beta 
+ \expvals{\hat X}_\beta^2$ \\
\hline

\makecell{Parameter controlling\\the limiting cases}
& Single mode frequency: $\omega$
& Surface gravity: $\kappa$\\
\hline

\end{tabular}}

\caption{analogue between black-body radiation and the quantum-gravity area model.
The classical variable $a$ denotes the averaged area density of the spherical
cross sections of the stretched horizon. In the null limit, where the stretched
horizon approaches the boundary of the causal diamond, this quantity becomes
the square of the causal-diamond length,
$a \xrightarrow{h_0 \to0} L^2$. Accordingly, in the same limit, the quantum
operator $\hX$ becomes the area operator: $\hat{A} = 4\pi \hat{X} + \mathcal{O}(h_0)$. The states $\ket{\alpha}$ are the
canonical Glauber coherent states, whereas $\ket{\gamma;a,b}$ are the affine
intelligent states constructed in Section~\ref{sec:intelligentstates}.}
\label{table:analogue}
\end{table}

The paper is organized as follows. First, we give a short review of Einstein fluctuation argument for black-body radiation in its modern quantum-optical form in Section \ref{sec:energyfluctuationinblackbodyradiation}. While this serves as a reminder, this also allows us to frame the argument in a convenient way for the construction of the gravitational analogue. Sections \ref{sec:classicaltheory} and \ref{sec:quantumtheory} develop respectively the classical and quantum versions of the gravitational theory. They contain the main technical ingredients of the construction, and may be skipped on a first reading without obscuring the conceptual argument. As such, Section \ref{sec:thermalstateandareafluctuation} starts with a convenient review of the technical results of the preceding sections, before delving into the construction of the gravitational thermal state and the associated area fluctuations. The implications of the results are discussed in Section \ref{sec:conclusion}, together with remarks on future directions. The appendices collect the technical details of the stretched-horizon construction, and review both the coadjoint-orbit description of the affine group and Einstein's original statistical derivation of the fluctuation formula.\par

\section{Energy Fluctuation in Black-Body Radiation}\label{sec:energyfluctuationinblackbodyradiation}
We start by presenting the modern treatment of the black-body radiation model. In doing so, we also highlight several aspects that will later prove important in constructing the gravitational analogue. Since the gravitational side of the correspondence involves a quantum system, the analogue is most naturally understood within the modern quantum-optical treatment of black-body radiation. We closely follow the treatment given in \cite{Gerry_Knight_2004}. Einstein’s original thermodynamic derivation is provided in Appendix \ref{app:statisticalderivation}. In this Section and for the rest of the paper, we write the expectation value of a quantum operator $\hat{U}$ in a quantum state $\ket{\psi}$ by
\begin{equation}
    \expvals{\hat U}_\psi = \mel{\psi}{\hat U}{\psi}.
\end{equation}
Additionally, the variance (or fluctuation squared) is denoted by
\begin{equation}
    \expvals{\Delta \hat U^2}_\psi = \expvals{\hat U^2}_\psi - \expvals{\hat U}_\psi^2.
\end{equation}\par
In order to model black-body radiation inside a cavity of volume $V$, we consider the single-mode electromagnetic field of frequency $\omega$ satisfying Maxwell's equations in Cartesian coordinates $x,y,z$ (in units where $\mu_0=1$, $\epsilon_0=1$, $c=1$)
\begin{align}
    E_x(z,t) &= \qty(\frac{2 \omega^2}{V})^{1/2} q(t)\sin(\omega z),\label{E}\\
    B_y(z,t) &= \qty(\frac{1}{\omega})\qty(\frac{2\omega^2}{V})^{1/2} \dot q(t) \cos(\omega z)\label{B},
\end{align}
where $q(t)$ is the amplitude of the field. The Hamiltonian of the system can be written
\begin{equation}
    H = \frac12 \int \dd V\, \qty[ E_x^2(z,t) + B_{y}^2(z,t)] = \frac12\qty(p^2 + \omega^2 q^2),
\end{equation}
where we denoted the momentum associated with the amplitude by $p = \dot q(t)$. Thus, the single mode electromagnetic field is a harmonic oscillator. The quantized theory is obtained by introducing the creation and annihilation operators such that
\begin{align}
    \hat{E}_{x}(z,t) &= \mathcal{E}_0 \qty(\hat a +\hat a^\dag)\sin(\omega z),\\
    \hat{B}_{y}(z,t) &= \mathcal{B}_0 \frac{1}{i}\qty(\hat a - \hat a^\dagger)\cos(\omega z),
\end{align}
where $\mathcal{E}_0 = \qty(\hbar \omega/ V)^\frac12$ and $\mathcal{B}_0 = \qty(1 /\omega)\qty( \hbar \omega^3/V)^\frac12$ can be understood respectively as the electric and magnetic fields ``per photon''. The creation and annihilation operators act on the Fock space in the standard way
\begin{align}
    a\ket{n} = \sqrt{n}\ket{n-1}\,,\quad a^\dag\ket{n} = \sqrt{n+1}\ket{n+1}
\end{align}
The Hamiltonian can be written as
\begin{equation}
    \hat H = \hbar \omega\qty(\hat a^\dag \hat a + \frac12) = \hbar\omega\qty(\hat{N}+\frac12),
\end{equation}
where we defined the number operator $\hat{N}$, counting the number of photons.\par
Let us now consider the coherent state
\begin{equation}\label{cancofstate}
    \ket{\alpha} = e^{-\frac{\abs{\alpha}^2}{2}}\sum_n \frac{\alpha^n}{\sqrt{n!}}\ket{n}.
\end{equation}
The absolute value of $\alpha$ is related to the amplitude of the fields 
\begin{equation}
    \expvals{\hat{E}_x(z,t)}_{\alpha} =  2\abs{\alpha} \mathcal{E}_0 \sin(\omega z)\cos\theta, 
\end{equation}
where $\alpha = \abs{\alpha}e^{i\theta}$. The modulus square of $\alpha$ is related to the expectation value of the Hamiltonian in the coherent state
\begin{equation}\label{eq:hamiltonianexpectationvaluealpha}
\expvals{\hat H_{\mathrm{rad}}}_{\alpha} = \hbar\omega  \abs{\alpha}^2,
\end{equation}
where we defined the radiative Hamiltonian by subtracting the vacuum energy
\begin{equation}
    \hat H_{\mathrm{rad}} = \hat H - \frac{\hbar\omega}{2}.
\end{equation}
Computing the squared fluctuation of the radiative Hamiltonian
\begin{equation}\label{eq:coherentstatehamiltonianfluctuation}
   \expvals{\Delta \hat{H}_{\mathrm{rad}}^2}_{\alpha}= \qty(\hbar\omega)^2 \abs{\alpha}^2,
\end{equation}
we obtain the fluctuation scaling for a coherent state
\begin{equation}\label{WZscaling}
     \expvals{\Delta \hat{H}_{\mathrm{rad}}^2}_{\alpha} =\hbar \omega \expvals{\hat H_{\mathrm{rad}}}_\alpha.
\end{equation}
This is a direct consequence of the probability distribution of the number operator in a coherent state. Indeed the probability to measure $n$ photons in a coherent state is given by
\begin{equation}
    P_n = \abs{\braket{n}{\alpha}}^2 = e^{-\abs{\alpha}^2}\frac{\abs{\alpha}^{2n}}{n!} = e^{-\langle \hat N \rangle_\alpha} \frac{\langle \hat N \rangle_\alpha^n}{n!},
\end{equation}
which is a Poisson distribution with a mean of $\langle \hat N \rangle_\alpha$, characteristic of the counting of independent discrete events. In this case we have the fluctuation scaling  $\expvals{\Delta \hat{U}^2} \sim \expvals{\hat U}$, which has the characteristic Brownian motion form. This behaviour, as we will see shortly, characterizes the high energy end, the Wien regime of the black-body spectrum. In the gravitational case, this scaling of the fluctuation characterizes the Verlinde--Zurek regime \cite{Verlinde:2019xfb,Zurek:2022snowmass}.\par
In the opposite low-energy Rayleigh--Jeans regime we have classical wave amplitudes that naturally fluctuate as a continuous variable instead. In order to see this behaviour, we can characterize the complete radiation of the black-body by a thermal state. First, note that the expectation values computed up until now can be understood as the quantum average in the state defined by the pure state density operator
\begin{equation}
    \rho_{\alpha} = \ket{\alpha}\bra{\alpha}.
\end{equation}
The thermal density operator at temperature $T = 1/\beta$ is given instead by
\begin{equation}\label{rhobeta}
    \rho_{\beta} = (1- e^{-\beta\hbar\omega})\sum_n e^{-\hbar\omega\beta n }\ket{n}\bra{n}.
\end{equation}
We can compute the average energy in the thermal state
\begin{equation}\label{eq:meanoccupationnumbethermal}
    \expvals{\hat{H}_{\mathrm{rad}}}_\beta \defeq \Tr(\rho_{\beta} \hat{H}_{\mathrm{rad}}) = \frac{\hbar \omega}{e^{\beta\hbar\omega}-1},
\end{equation}
which is the Bose-Einstein distribution. Computing the squared fluctuation of the radiative Hamiltonian gives the formula in Table \ref{table:analogue}
\begin{equation}\label{eq:energythermal}
    \expvals{\Delta \hat{H}_{\mathrm{rad}}^2}_\beta = \hbar \omega \expvals{\hat{H}_{\mathrm{rad}}}_{\beta} + \expvals{\hat{H}_{\mathrm{rad}}}_\beta^2.
\end{equation}
For the thermal state, we therefore obtain both wave-like scaling and particle-like scaling of the fluctuations.
The limiting cases of the Wien and Rayleigh--Jeans laws can be deduced directly
from equations \eqref{eq:meanoccupationnumbethermal} and
\eqref{eq:energythermal}. In the low-frequency, or Rayleigh--Jeans, regime
\(\beta\hbar\omega \ll 1\), one has
\begin{equation}\label{eq:RJregime}
    \expvals{\hat H_{\mathrm{rad}}}_\beta
    \simeq
    \frac{1}{\beta}.
\end{equation}
The linear term in \(\expvals{\hat H_{\mathrm{rad}}}_\beta\) is then negligible
compared to the quadratic term, and Einstein's fluctuation formula reduces to
\begin{equation}
     \expvals{\Delta \hat{H}_{\mathrm{rad}}^2}_\beta
     \simeq
     \expvals{\hat H_{\mathrm{rad}}}_\beta^2,\qquad  \hbar\omega
    \ll \expvals{\hat H_{\mathrm{rad}}}_\beta.
\end{equation}
This is the wave-like contribution to the energy fluctuations. To see this consider the classical electromagnetic standing wave (Equations \eqref{E} and \eqref{B}) with the classical Boltzmann distribution $P(E) = \beta e^{-\beta E}$. Then we find
\begin{align}\label{averaEE2}
    \overline{E} =\beta \int_0^\infty dE\,E\, e^{-\beta E} = \frac1\beta \,,\quad \overline{E^2} =\beta \int_0^\infty dE\,E^2 e^{-\beta E} = \frac2{\beta^2} 
\end{align}
and therefore
\begin{align}
  \Delta  \overline{E}^2=  \beta^{-2} = \overline{E}^2
\end{align}
On the other hand, in the high-frequency, or Wien, regime
\(\beta\hbar\omega \gg 1\), one has
\begin{equation}
    \expvals{\hat H_{\mathrm{rad}}}_\beta
    \simeq
    \hbar\omega e^{-\beta\hbar\omega}. 
\end{equation}
In that regime, the quadratic term is negligible compared to the linear term and
Einstein's fluctuation formula then becomes
\begin{equation}
     \expvals{\Delta \hat{H}_{\mathrm{rad}}^2}_\beta
     \simeq
     \hbar\omega \expvals{\hat H_{\mathrm{rad}}}_\beta,\qquad  \hbar\omega
    \gg \expvals{\hat H_{\mathrm{rad}}}_\beta
\end{equation}
This is the particle-like contribution to the energy fluctuations. Looking at \eqref{rhobeta} we see that, in the Wien regime, the states with $n>0$ are strongly suppressed, and we have to do with dramatic depletion of high energy states. Curiously, this behaviour also appears in the model of Cohen, Kaplan and Nelson \cite{Cohen:1998zx} which aims at incorporating some quantum gravity effects to particle physics. The depletion of UV degrees of freedom in the CKN model is discussed in \cite{Banks:2019arz}.
\par
For the analogueous construction in the gravitational setting, it will prove convenient to write the thermal state in the coherent states basis. The main observation is that the thermal state admits a decomposition in the coherent states basis with a Gaussian Glauber-Sudarshan function \cite{Sudarshan:1963ts,Glauber:1963tx}
\begin{equation}
    \rho_\beta = \frac{\hbar \omega}{\pi \expvals{\hat{H}_{\mathrm{rad}}}_\beta}\int_{\mathbb{C}} \dd^2 \alpha\,  e^{-\frac{\hbar\omega}{\expvals{\hat H_{\mathrm{rad}}}_\beta} \abs{\alpha}^2} \ket{\alpha}\bra{\alpha}.
\end{equation}
The remarkable property of coherent states is that they provide the bridge between the quantum and classical observables. Indeed, expectation values of quantum operators in these states reproduces the classical values\footnote{This property can be generalized, and was used in the context of the corner proposal to relate abstract representation theory of the corner symmetry groups with classical geometrical observables like the area \cite{Ciambelli:2025ztm,Kowalski-Glikman:2025hom,Varrin:2025okc,Varrin:2026bmb}.}. We can therefore
use equation \eqref{eq:hamiltonianexpectationvaluealpha} to rewrite the quantum thermal state in terms of a distribution over the classical energy $H_{\mathrm{rad}}$
\begin{equation}\label{eq:citethisone}
    P_\beta(H_{\mathrm{rad}}) = \frac{1}{\expvals{\hat H_{\mathrm{rad}}}_\beta}e^{-\frac{H_{\mathrm{rad}}}{\expvals{\hat H_{\mathrm{rad}}}_\beta}},
\end{equation}
which is the distribution given in Table \ref{table:analogue}. Note that the form of this distribution is fixed by the defining property that thermal states maximize the Shannon entropy of the probability distribution, under the constraint that the average of the operator in the thermal state is fixed to the classical value \cite{Jaynes:1957zza}. Of course, in the Rayleigh--Jeans limit \eqref{eq:RJregime}, we recover the standard Boltzmann distribution
\begin{equation}
  P_\beta(H_{\mathrm{rad}}) \approx  \beta e^{-\beta H_{\mathrm{rad}}}, \qquad  \hbar\omega
    \ll \expvals{\hat H_{\mathrm{rad}}}_\beta
\end{equation}
which describes a classical thermal distribution of electromagnetic standing waves. \par
The emergence of both scaling terms in the thermal state can  be better understood using the law of total variance, which states that, for a density operator written as a classical mixture, the variance separates into an average intra-state contribution, measuring the quantum fluctuations within each component of the mixture, and an inter-state contribution, measuring the classical fluctuations of the conditional expectation values across the ensemble. In the case of the black-body thermal state the total variance law gives
\begin{equation}\label{eq:totalvariance}
    \expvals{\Delta \hat{E}^2}_\beta
    =
    \int_0^\infty \dd E\, P_\beta(E)
    \expvals{\Delta \hat{E}^2}_{\alpha(E)}
    +
    \int_0^\infty \dd E\, P_\beta(E)
    \qty(E - \expvals{\hat E}_{\beta})^2 .
\end{equation}
where we denoted $H_{\mathrm{rad}} = E$ for simplicity and where $ \expvals{\Delta \hat{E}^2}_{\alpha(E)}$ is given by equation \eqref{eq:coherentstatehamiltonianfluctuation} after identifying $\hbar \omega\abs{\alpha}^2 = E$. Thus the particle-like Wien contribution comes from the intrinsic Poissonian fluctuations within each coherent state, whereas the wave-like Rayleigh--Jeans contribution comes from the classical thermal spread of the energy.

\par
This rewriting of the thermal state is particularly useful, as it does not require the full knowledge of the Hamiltonian. Since this procedure will serve as the guiding principle in the construction of the gravitational analogue, we give here a summary of the relevant steps. First, we identify a family of states whose expectation value reproduces the classical value of the observable and whose fluctuations obey the Wien scaling. Next, we define a probability distribution over the classical values by requiring that it maximizes its Shannon entropy under the constraint of a fixed thermal average for the Hamiltonian. Finally, the thermal density operator is obtained by forming the corresponding mixture of these states, namely by integrating the projectors onto the states against the entropy-maximizing probability distribution.
\section{Classical Gravitational Phase Space}\label{sec:classicaltheory}
In order to discuss the gravitational analogue of the black-body radiation, we consider one of the simplest non-trivial gravitational systems: a causal diamond in a Minkowski spacetime. The thermodynamical properties of causal diamonds and their relationship with quantum gravity have been investigated for a long time \cite{Martinetti:2002sz,Jacobson:2015hqa,Jacobson:2018ahi,
Banks:2020zcr,Arzano:2020thh,Chakraborty:2022qdr,Jacobson:2022gmo,Jacobson:2022jir,Fransen:2025npa}. The boundary of the causal diamond is a null hypersurface. The standard method to sidestep the complications that come with a null structure is to consider a timelike \textit{stretched horizon} that deforms into the null boundary of the causal diamond in some well-defined limit. It often proves easier to perform the symplectic analysis on the stretched horizon first and take the null limit at the end. 
This exact procedure was performed in \cite{Ciambelli:2025flo} in the context of a causal diamond in Minkowski spacetime. While we will give a brief review of the main results shortly, the relevant observation for the present work is that the averaged area density over the stretched horizon, together with its derivative away from the horizon form an affine algebra. More details of the construction are presented in Appendix \ref{app:stretchedhorizon}, and we refer the reader to \cite{Price:1986yy,Thorne:1986iy} (see also \cite{Freidel:2024gch} for a modern treatment) for a more thorough description of the stretched horizon.

\begin{figure}
        \centering
        \includegraphics[width=0.6\linewidth]{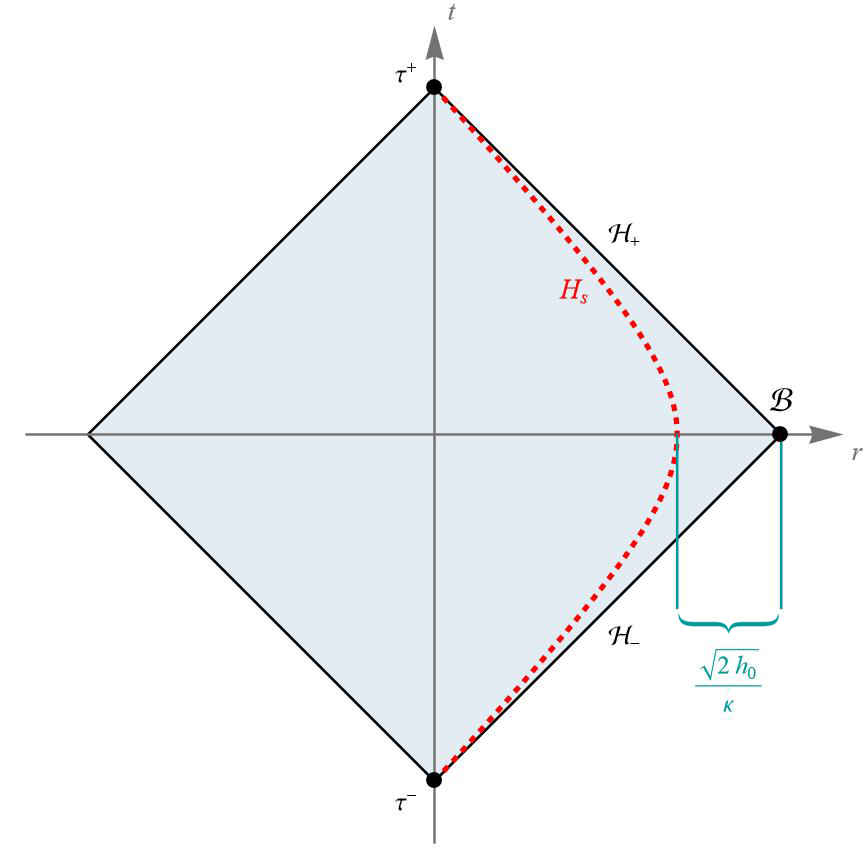}
        \caption{The causal diamond and stretched horizon $H_s$. The bifurcating surface situated at $t=0, r=L$ is denoted by $\mathcal{B}$, while the caustics of the stretched horizons are denoted by $\tau^\pm$. The future and past horizons are denoted by $\mathcal{H}_\pm$. The parameter $h_0$ controls how much the stretched horizon differs from the causal diamond's boundary. In particular, at $t=0$, the radial distance between the stretched horizon and the bifurcating sphere is given by $\sqrt{2h_0}/\kappa$.}
        \label{fig:causaldiamond}
    \end{figure}

We consider a stretched horizon $H_s$ inside a finite causal diamond in four-dimensional Minkowski spacetime, see Fig.~\ref{fig:causaldiamond}.    
The line element can be written in Gaussian-null coordinates adapted to
$H_s$
\begin{equation}
\dd s^2
=-2(h_0 + \kappa \rho)\dd \tau^2  + 2 \dd \tau \dd\rho + 
\varphi(\tau,\rho)d\Omega_2^2,
\end{equation}
with
\begin{equation}\label{eq:areadensityfield}
\varphi(\tau,\rho)
=
\left(
L
-
\frac{1}{2\kappa}e^{\kappa\tau+\alpha}
-
\left(\rho+\frac{h_0}{\kappa}\right)e^{-\kappa\tau-\alpha}
\right)^2 .
\end{equation}
Here $L$ is the radius of the bifurcation two-sphere, so that its area is given by $A=4\pi L^2$. The constant $\kappa$ is the inaffinity of the
clock vector $\ell=\partial_\tau$, while $h_0$ is a dimensionless parameter
fixing the location of the stretched horizon relative to the null horizon. The stretched horizon is located at $\rho=0$, and is
parametrized by a clock coordinate $\tau$ and angular coordinates
$\vartheta^A$ on $S^2$. Its endpoints occur at $\tau=\tau_\pm$, where the
spherical cross-section shrinks to a caustic, and we denote
\begin{equation}
T=\tau_+-\tau_- .
\end{equation}
The constant $\alpha$
is an integration constant corresponding to a shift of the origin of $\tau$.
\par 
The field $\varphi(\tau,\rho)$ encodes the transverse area density of the
spherical cross-sections. In the stretched horizon literature, it is sometimes called the \textit{breathing mode} \cite{Binetruy:1993sfbd,Seraj:2021bmd,Li:2022geo,Bub:2023qgb,Lee:2023aib,Vermeulen:2024gquest}.
The field
\begin{equation}
  \psi(\tau)
  =
  \left.\partial_\rho \varphi(\tau,\rho)\right|_{\rho=0}
\end{equation}
encodes the first-order change of this transverse area density as one
moves away from the stretched horizon in the $\rho$ direction.
Next, we take the average of these fields over the stretched horizon time
\begin{equation}\label{eq:spin0fields}
\bar{\varphi}
=
\frac{1}{ T}
\int_{\tau_-}^{\tau_+}d\tau\,
\varphi(\tau),
\qquad
\bar{\psi}
=
\frac{1}{T}
\int_{\tau_-}^{\tau_+}d\tau
\,
\psi(\tau).
\end{equation}
As is shown in Appendix \ref{app:stretchedhorizon}, the null limit of the averaged field $\varphi$ corresponds to the area of the causal diamond's boundary
\begin{equation}
    \lim_{h_0\to 0}\bar\varphi = \frac{A}{4\pi}.
\end{equation}
The symplectic analysis of the stretched horizon gives the following Poisson bracket for the averaged fields \cite{Ciambelli:2025flo}
\begin{equation}\label{eq:poissonbracket}
\{\bar{\psi},\bar{\varphi}\}
=
\frac{G}{h_0T}\bar{\varphi}.
\end{equation}
Thus, the averaged fields realize an affine algebra in the covariant phase space. This formula is the starting point of our investigation.\par
The strategy we will pursue is to quantize the above bracket in order to compute fluctuations of the averaged area density field $\bar\varphi$. In the $h_0 \to 0$ limit, this will then correspond to the fluctuations of the causal diamond's area operator. In order to keep track of the $h_0 \to 0$ limit, we need to understand clearly how the quantum operators map to classical observables. In the context of quantum systems with symmetries, this is usually done by relating coherent states with coadjoint orbits. We therefore start by describing the classical phase space with Poisson bracket \eqref{eq:poissonbracket} in terms of coadjoint orbits of the affine group.\par
The two-dimensional real affine algebra $\mathfrak{aff}_+(\RR)$ is generated by the basis $\qty{X,D}$ with Lie commutation relation
\begin{equation}\label{eq:affinealgebra}
    \qty[D,X] = -X.
\end{equation}
Since the area density is always positive, the coadjoint orbit that can be associated with the averaged gravitational phase space is the upper half plane
\begin{equation}
    \mathcal{O}_{\mathrm{Aff}}^+ = \qty{(x,d)\mid x>0,d\in \RR}.
\end{equation}
The Kirillov--Kostant--Sourieau two-form on that orbit induces the following Poisson bracket on the coordinates
\begin{equation}
    \qty{d,x} = -x.
\end{equation}
Comparing this to the bracket \eqref{eq:poissonbracket}, we obtain the following identifications of the averaged fields in terms of coadjoint orbit coordinates
\begin{equation}\label{eq:fieldcoordinatesidentification}
    \bar\varphi = x, \qquad \bar\psi = -\frac{G}{h_0 T}d.
\end{equation}
Below we will find it convenient to introduce Darboux coordinates on the orbits $Q=x,P= \frac{d}{x}$, which have canonical bracket
\begin{equation}\label{QP}
    \qty{Q,P} = 1.
\end{equation}
For more details on the coadjoint orbits of the affine group, we refer to Appendix \ref{app:coadjointorbits}.
\section{Quantum Theory}\label{sec:quantumtheory}
We are now ready to introduce the quantum theory associated with the classical gravitational phase space described in the previous section. First, we will only be interested in the phase space consisting of the averaged fields $\bar\varphi$ and $\bar\psi$, which is what allows us to have full control over the quantum theory. It is important, however, to emphasize that this simplified setup, while sufficient for studying area fluctuations, is not a complete theory of quantum gravity. We only retain the zero mode of the stretched-horizon time dependence of the fields, and we neglect quantum fluctuations in the angular directions by fixing the background to Minkowski spacetime. The lack of a complete theory is precisely the reason why we resort to statistical arguments to discuss the quantum aspect of the area operator. In that sense, our agnostic position with respect to quantum gravity is not dissimilar to the one early 20th century physicists had with respect to light.\par
Since the classical theory of the averaged fields is described by coadjoint orbits of the affine group, the corresponding quantum theory is naturally obtained by applying Kirillov’s orbit method, namely by geometrically quantizing the affine coadjoint orbits \cite{Gayral:2007zza}. As expected, this procedure simply yields the standard representation theory for the affine group originally described by Gelfand and Naimark \cite{GelfandNaimark1947}. These representations will therefore be our starting point for the quantum theory. We also discuss the associated uncertainty relations, which will play a central role in the construction of the thermal state. Next, we construct the affine coherent states in order to connect the representation theory to the classical variables described in the previous section. Finally, we construct states that minimize the Heisenberg--Robertson uncertainty relations, called \textit{intelligent states}. These will be the building blocks of the gravitational thermal state.
\par
Consider the affine group associated with the Lie algebra \eqref{eq:affinealgebra}
\begin{equation}
    \mathrm{Aff}_+(\RR) = \qty{(q,p) \in (0,\infty)\times \RR \mid (q,p)\cdot (q',p') = (qq',p+\frac{p'}{q})}.
\end{equation} 
The irreducible unitary representations of this group are all unitarily equivalent to  
\begin{equation}
    U_{q,p}: L^{2}(\RR_+,\dd x)\rightarrow L^2(\RR_+,\dd x), \quad  (U_{q,p}\psi)(x) = \frac{e^{ipx}}{\sqrt{q}}\psi\qty(\frac{x}{q}).
\end{equation}
The associated quantum generators in the affine algebra act as differential operators
\begin{equation}\label{eq:Ddifferentialoperato}
    \hX \psi(x) = x \psi(x), \quad \hD \psi(x) = -i\qty(x\partial_x + \frac12)\psi(x).
\end{equation}
For any state, and any two operators $\hat A$, $\hat B$ the Schrödinger-Robertson inequality for the affine algebra reads
\begin{equation}\label{eq:robertstonschrodinger}
    \expvals{\Delta \hat A^2}\expvals{\Delta \hat B^2} \geq \frac14 \abs{\expvals{[\hat A,\hat B]}}^2 + \sigma^2_{AB},
\end{equation}
where $\sigma_{AB}$ is the covariance between the two operators, defined as
\begin{equation}
    \sigma_{AB}\defeq \frac12 \expvals{\hat{A}\hat{B}+\hat{B}\hat{A}} - \expvals{\hat{A}}\expvals{\hat{B}}.
\end{equation}
If the state is such that the covariance vanishes, we obtain the standard Heisenberg--Robertson relation
\begin{equation}\label{eq:heisenberg}
    \sqrt{\expvals{\Delta \hat{A}^2} \expvals{\Delta \hat{B}^2}} \geq \frac12 \abs{\expvals{\qty[\hat{A},\hat{B}]}}.
\end{equation}
\subsection{Affine coherent states}
In order to connect the representations to the classical theory described in Section \ref{sec:classicaltheory}, we consider Perelomov affine coherent states. Since the stabilizer for the affine group is trivial, these states can be constructed by displacing a fiducial state $\phi_0(x)$ by the action of the group
\begin{equation}\label{eq:displacementformula}
    \phi_{p,q}(x) = \qty(U_{p,q} \phi_0)(x) = \frac{e^{ipx}}{\sqrt{q}}\phi_0\qty(\frac{x}{q}).
\end{equation}
In the present work, it will be convenient to work with the following fiducial state
\begin{equation}
    \phi^\nu_0(x) = \frac{2^\nu}{\sqrt{\Gamma\qty(2 \nu)}} e^{-x}x^{\nu-\frac12}, 
\end{equation}
where $\nu>\frac12$. Using the general formula \eqref{eq:displacementformula}, the Perelomov coherent states are then written
\begin{equation}\label{eq:perelomovaffinecoherentstates}
    \phi^\nu_{q,p}(x) = \frac{e^{ipx}}{\sqrt{\Gamma(2\nu)}}\qty(\frac{2}{q})^\nu e^{-\frac{x}{q}} x^{\nu-\frac12}.
\end{equation}
These states are normalized in the Hilbert space $L^2(\RR_+,\dd x)$ and 
they form a closure of the identity
\begin{equation}
    \frac{2\nu-1}{2}\int_0^\infty \int_{-\infty}^\infty \frac{\dd q}{2\pi}\dd p\, \ket{q,p}\bra{q,p} = \mathds{1},
\end{equation}
where we used the abstract $\hat{X}$ eigenstates $\ket{x}$ to define
\begin{equation}
    \ket{q,p} = \int_0^\infty \dd x\, \phi_{p,q}(x)\ket{x}.
\end{equation}
Using the differential representation of the operators \eqref{eq:Ddifferentialoperato}, we can compute
\begin{equation}\label{eq:coherentstateexpectationvalues}
    \expvals{\hX}_{q,p} =  \nu q , \quad \expvals{\hD}_{q,p} = \nu qp.
\end{equation}
Therefore, the Perelomov coherent state $\psi_{q,p}^\nu(x)$ is associated with the point $Q= \nu q, P=  p$ on the coadjoint orbit (see Equation \eqref{QP}). Using the identification between the fields and the coadjoint orbit coordinates \eqref{eq:fieldcoordinatesidentification}, we can identify the parameters of the coherent states with the physical fields as
\begin{equation}\label{eq:momentmapindentification}
    \nu q = \bar\varphi, \qquad \nu pq = -\frac{h_0 T}{G}\bar\psi.
\end{equation}
\par
To conclude our discussion of affine coherent states, we note that the fluctuation of the $\hX$ operator is given by
\begin{equation}
    (\Delta \hX)^2_{\nu,q,p} = \frac{\nu}{2}q^2,
\end{equation}
which corresponds to the wave-scaling behaviour of the low energy Rayleigh--Jeans law. The affine coherent states analyzed here are therefore not the right analogue of the canonical coherent state  \eqref{cancofstate} in the black-body case, where it is associated with the high-energy Wien regime. This is because the states \eqref{eq:perelomovaffinecoherentstates} do not saturate the Heisenberg--Robertson uncertainty relation \eqref{eq:heisenberg}. As we will see in the next section, the states that do saturate the uncertainty relation have the appropriate particle-scaling needed to construct the thermal state.
\subsection{Intelligent states}\label{sec:intelligentstates}
\begin{figure}[H]
        \centering
        \includegraphics[width=0.7\linewidth]{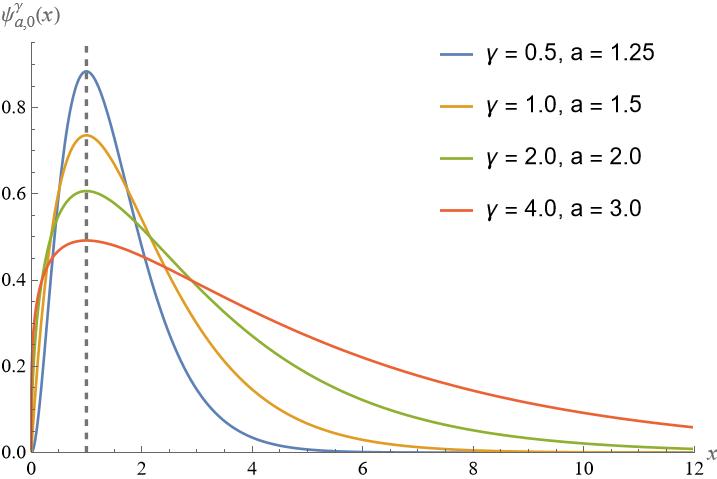}
        \caption{Affine intelligent states $\psi^\gamma_{a,0}(x)$ for $b=0$ and for parameters satisfying
$a-\gamma/2=1$. The dashed line marks the common maximum of the distributions. The
parameter $\gamma$ controls the intrinsic spread of the state: smaller $\gamma$ gives a
more sharply localized area-density state, while larger $\gamma$ gives a broader,
asymmetric distribution on the positive half-line.}
        \label{fig:placeholder}
\end{figure}

States that saturate the operators $\hat A$ and $\hat B$ uncertainty relation \eqref{eq:robertstonschrodinger} are called \textit{intelligent states} \cite{jackiw1968,Brif:1997kp} and can be obtained by the eigenvalue equation
\begin{equation}
    \qty(\eta \hat{A} + i \hat{B})\ket{\lambda,\eta} = \lambda \ket{\lambda,\eta},
\end{equation}
where $\eta$ is an arbitrary complex number labeling the state and $\lambda$ is a complex eigenvalue. In the case where $\eta$ is real, the eigenvalue equation can be written
\begin{equation}\label{eq:intelligenteigenvalueequation}
    \qty(\hat{A}+i\gamma \hat{B})\ket{\lambda,\gamma} = \lambda\ket{\lambda,\gamma},
\end{equation}
where $\gamma$ is a real parameter. Solutions to the above equation saturate the Heisenberg relation \eqref{eq:heisenberg}.
For the affine group, we can construct the wave function of such a state using the differential operator representation
\begin{equation}
  \qty[ x + \gamma \qty(x\partial_x+\frac12)] \psi^\gamma_\lambda(x) = \lambda \psi^{\gamma}_{\lambda}(x).
\end{equation}
The above equation is solved by
\begin{equation}
    \psi_\lambda^\gamma(x) = C e^{-\frac{x}{\gamma}} x^{\frac{2\lambda-\gamma}{2\gamma}}.
\end{equation}
where $C$ is a constant.
For $\gamma>0$ and $\Re(\lambda)>0$, the states are normalizable
\begin{equation}
    \int_0^\infty \dd x\, \abs{\psi_\lambda^\gamma(x)}^2 =\abs{C}^2 \int_0^\infty \dd x\, e^{-2\frac{x}{\gamma}}x^{\frac{2\Re(\lambda)}{\gamma}-1}
     = \abs{C}^2 \qty(\frac{\gamma}{2})^{2\frac{\Re\lambda}{\gamma}} \Gamma\qty(\frac{2\Re\lambda}{\gamma}).
\end{equation}
Thus the normalized intelligent states are given by
\begin{equation}
    \psi_{a,b}^\gamma(x) = \qty(\frac{2}{\gamma})^{\frac{a}{\gamma}}\Gamma\qty(\frac{2a}{\gamma})^{-\frac12} e^{-\frac{x}{\gamma}}x^{\frac{a}{\gamma}-\frac12}x^{i\frac{b}{\gamma}}, \quad a>0,\gamma>0, b\in\RR.
\end{equation}
where we denoted $\operatorname{Re}\lambda=a, \operatorname{Im}\lambda=b$. The intelligent states are therefore distributions in $x$ with maximum at $x_{max} = a- \gamma/2$ and with width controlled by $\gamma$ (see Fig.\ \ref{fig:placeholder})

We can compute the expectation values of the operators in these states 
\begin{align}
    \expvals{\hX}_{a,b,\gamma} &=  a,\label{eq:Xexpvalintelligent}\\
    \expvals{\hD}_{a,b,\gamma} &= \frac{b}{\gamma}.
\end{align}
as well as the squared fluctuation
\begin{align}
    \expvals{\Delta \hX^2}_{a,b,\gamma} &= \frac{\gamma}{2} a,\\
     \expvals{\Delta \hD^2}_{a,b,\gamma} &= \frac{a}{2\gamma},
\end{align}
which indeed saturates the Robertson equality
\begin{equation}
  \sqrt{\expvals{\Delta \hX^2}_{a,b,\gamma}\expvals{\Delta \hD^2}_{a,b,\gamma}} =\frac12 a.
\end{equation}

\par
Contrary to the Heisenberg structure of the black-body case, the intelligent and affine coherent states of the affine algebra do not coincide. They are however related by a $x$-dependent phase
\begin{equation}
    \psi^\gamma_{a,b}(x) = e^{i\qty(\frac{b}{\gamma}\ln x - px)}\phi^{\frac{a}{\gamma}}_{\gamma,p}(x).
\end{equation}
The two families of states therefore have the same probability density in the
\(X\)-representation.\par
Using the moment-map identification \eqref{eq:momentmapindentification}, the
Perelomov coherent states associate the classical averaged fields
$(\bar\varphi,\bar\psi)$ with points on the affine coadjoint orbit. Since the
intelligent states differ from the Perelomov coherent states only by an
$x$-dependent phase, all expectation values of
operators constructed solely from $\hat X$ coincide in the two families of
states. In particular
\begin{equation}
    \expvals{\hat X}_{a,b,\gamma} = a,
\end{equation}
reproduces the classical averaged area density
\begin{equation}
    a = \bar\varphi .
\end{equation}
Importantly, the identification of $a$ with the classical area density is
independent of the stretched-horizon parameter $h_0$. The reason is that $h_0$
only enters the identification of the conjugate observable $\bar\psi$ (c.f. Equation \eqref{eq:fieldcoordinatesidentification})
whereas the area-density observable itself is directly identified with the
affine coadjoint orbit coordinate $Q$. Consequently, all fluctuations associated with $\hat X$
are insensitive to the precise location of the stretched horizon and admit a
smooth null-horizon limit $h_0 \to 0$.
This is crucial for the interpretation of $\hat X$ as the causal-diamond area
operator in the null limit.
\section{Thermal State and Area Fluctuation}\label{sec:thermalstateandareafluctuation}
Before diving into the construction of the gravitational thermal state, we give here a review of the most relevant results of Section \ref{sec:classicaltheory} and \ref{sec:quantumtheory}. The classical gravitational system considered is a stretched horizon inside a causal diamond in Minkowski spacetime. This is a timelike surface, characterized by a dimensionless parameter $h_0$ which measures how far the stretched horizon is from the null boundary of the causal diamond. Within the Poisson algebra of observables on the stretched-horizon phase space,
the averaged modes form a finite-dimensional closed subalgebra realizing the
affine algebra
\begin{equation}
    \qty{\bar\varphi,\bar\psi}
    =
    -\frac{G}{h_0 T}\bar\varphi.
\end{equation}
Here $G$ is Newton's constant, $T$ is the stretched-horizon time interval between
the two caustics, $\bar\varphi$ is the time-averaged area density of the spherical
cross sections, and $\bar\psi$ measures the corresponding time-averaged variation
of this area density away from the stretched horizon. In the null limit $h_0 \to 0$, as the stretched horizon approaches the boundary of the causal diamond, the observable $\bar\varphi$ becomes its area.
In the quantum theory, the observables get promoted to operators acting on the representation space of the affine group
\begin{equation}
    \bar\varphi \rightarrow \hat{X}, \qquad -\frac{h_0 T}{G} \bar\psi \rightarrow \hD. 
\end{equation}
Finally, one can construct a family of states which reproduce the classical value of the area density operator when taking expectation values and which gives a Wien like particle-scaling of the fluctuations. We remind the reader that those are the exact properties needed for the procedure described at the end of section \ref{sec:energyfluctuationinblackbodyradiation}. These states are called \textit{intelligent} because they saturate the Heisenberg--Robertson uncertainty relations \eqref{eq:heisenberg} and are given by
\begin{equation}
     \psi_{a,b}^\gamma(x) = \qty(\frac{2}{\gamma})^{\frac{a}{\gamma}}\Gamma\qty(\frac{2a}{\gamma})^{-\frac12} e^{-\frac{x}{\gamma}}x^{\frac{a}{\gamma}-\frac12}x^{i\frac{b}{\gamma}}, \quad a>0,\gamma>0, b\in\RR.
\end{equation}
These states are normalized in the Hilbert space $L^2(\RR_+,\dd x)$.
Here, $a+ib$ is the complex eigenvalue of equation
\eqref{eq:intelligenteigenvalueequation}, playing the role of the
coherent-state label $\alpha \in \mathbb{C}$ in the canonical construction.
The parameter $\gamma$ plays the role of a fixed structural parameter analogueous
to $\hbar$. The expectation values of the $\hX$ operator and its squared fluctuation in the intelligent states are given by
\begin{equation}
    \expvals{\hX}_{a,b,\gamma} =  a, \quad \expvals{\Delta \hX^2}_{a,b,\gamma} = \frac{\gamma}{2} a,
\end{equation}
which indeed gives a Wien like scaling of the squared fluctuation
\begin{equation}
     \expvals{\Delta \hX^2}_{a,b,\gamma} = \frac{\gamma}{2} \expvals{\hX}_{a,b,\gamma}.
\end{equation}
In the context of area fluctuations, we call this the Verlinde--Zurek scaling \cite{Verlinde:2019ade,Verlinde:2019xfb,Banks:2021jwj,Zurek:2020vac,Verlinde:2022hhs,Ciambelli:2025flo}.
Finally, we were able to directly relate the parameter $a$ to the classical value of the averaged density area $\bar\varphi$.\par
Following the procedure described at the end of Section \ref{sec:energyfluctuationinblackbodyradiation}, we want to construct a thermal state by integrating an exponential distribution over the classical values of the area density.
Before we do so, let us note that it can be shown that the inaffinity of the clock vector of the stretched horizon becomes the surface gravity of the causal diamond's boundary in the $h_0 \to 0$ limit (c.f. Appendix \ref{app:stretchedhorizon}). At fixed surface gravity $\kappa$, the first law of horizon thermodynamics \cite{Bardeen:1973gs} suggests identifying the variation of the horizon energy with the variation of the causal-diamond area 
\begin{equation}
    \delta E_{\mathrm{hor}} = \frac{\kappa}{8\pi G}\delta A.
\end{equation}
Since, in the null limit, the averaged area density satisfies $A=4\pi \bar\varphi$, integrating the first law at fixed $\kappa$ motivates the effective identification
\begin{equation}\label{eq:classicalhorizonenergy}
    E_{\mathrm{hor}}(\bar\varphi) = \frac{\kappa}{2G}\bar\varphi + \mathcal{O}(h_0).
\end{equation}
Here, the additive integration constant has been set to zero. Although this relation is motivated thermodynamically rather than derived from a microscopic Hamiltonian, it is sufficient for constructing the maximum-entropy thermal state associated with area fluctuations.
In the quantum theory, the effective horizon Hamiltonian can be written
\begin{equation}
    \hat E_{\mathrm{hor}}= \frac{\kappa}{2G}\hat{X} + \mathcal{O}(h_0).
\end{equation}
Note that taking the expectation value of this operator in the intelligent state gives
\begin{equation}\label{eq:classicalhamiltonian}
    E_{\mathrm{hor}}(a) = \expvals{\hat E_{\mathrm{hor}}}_{\gamma,a,b}= \frac{\kappa}{2G}a + \mathcal{O}(h_0),
\end{equation}
which, with the identification \eqref{eq:momentmapindentification}, reproduces the classical value \eqref{eq:classicalhorizonenergy}. As we mentioned at the end of Section \ref{sec:energyfluctuationinblackbodyradiation}, this is a crucial part of the procedure to construct a thermal state.

Knowing what the energy is, we can now construct the statistical mechanics of areas, aiming at getting some insight into the quantum structure of causal diamonds. To this end we introduce a classical probability density over the averaged area density $P(a)$ \cite{Jaynes:1957zza}. In order to determine the distribution associated with a thermal state, we maximize the Shannon entropy $S[P]
=
-\int_0^\infty da\,P(a)\ln P(a)$ under the two constraints that the distribution be normalized 
\begin{equation}
    \int_0^\infty \dd a\, P(a) = 1,
\end{equation}
and that the classical average of the energy
\begin{equation}
    \bar{E}_\mathrm{hor} \defeq  \int_0^\infty \dd a\, P(a)E_{\mathrm{hor}}(a) 
\end{equation}
is fixed.
In the above, we have interpreted the expectation value of the area Hamiltonian as its classical value. As the Hamiltonian only depends on the operator $\hX$, its expectation value coincides with the one in the Perelomov coherent states, thereby justifying this interpretation.
Note that, using equation \eqref{eq:classicalhamiltonian}, the classical average energy is related to the classical average area
\begin{equation}
    \bar{a} \defeq \int \dd a\, a P(a) = \frac{2G}{\kappa}\bar{E}_{\mathrm{hor}} + \mathcal{O}(h_0).
\end{equation}
Introducing Lagrange multipliers $\lambda$ and $\beta$, the thermal distribution extremizes
\begin{equation}
    \mathcal F [P] = -\int_0^\infty \dd a\, P(a)\ln(P(a))- \lambda \qty(\int_0^\infty \dd a\, P(a)-1)-\beta \qty(\int_0^\infty\dd a\, P(a)E_{\mathrm{hor}}(a) - \bar{E}_\mathrm{hor}).
\end{equation}
This is easily solved by
\begin{equation}
    P_\beta(a) = \frac{\beta\kappa}{2G}e^{-\frac{\beta\kappa}{2G}a},
\end{equation}
where we used the normalization condition to fix the value of the Lagrange multiplier $\lambda$. The remaining Lagrange multiplier $\beta$ becomes the temperature of the thermal state in the familiar way \cite{Jaynes:1957zza}.\par
Now that we have the distribution, we can compute
\begin{equation}\label{eq:averageX}
    \bar{a} = \expvals{\hX}_\beta = \frac{2G}{\beta\kappa},
\end{equation}
which puts the distribution in an analogueous form to the black-body one \eqref{eq:citethisone}
\begin{equation}
    P_\beta(a) = \frac{1}{\expvals{\hX}_\beta} e^{-\frac{a}{\expvals{\hX}_\beta}}
\end{equation}
In analogue to what we did in Section \ref{sec:energyfluctuationinblackbodyradiation} in the black-body case, we now construct the thermal state through the intelligent states\footnote{{Strictly speaking, this thermal density operator still depends on the intelligent-state parameter $b$. This dependence affects expectation values of operators involving $\hat D$, but drops out for observables constructed solely from $\hat X$. Since the present work only uses the thermal state to compute area fluctuations, we keep $b$ fixed and suppress this dependence in what follows.}}

\begin{equation}
    \rho_{\beta} = \int_{0}^\infty \dd a\, P_\beta(a) \ket{\gamma;a,b} \bra{\gamma;a,b},
\end{equation}
where we defined
\begin{equation}
    \ket{\gamma;a,b} = \int_0^\infty \dd x\, \psi^\gamma_{a,b}(x) \ket{x}.
\end{equation}
The normalization of the thermal density operator directly follows from the normalization of the intelligent states and the one of the classical distribution.\par
We can now compute the squared fluctuation of the area density operator
\begin{equation}
    \expvals{\Delta \hat{X}^2}_\beta = \frac{4 G^2}{\beta^2\kappa^2} + \frac{G\gamma}{\beta\kappa}.
\end{equation}
 Using equation \eqref{eq:averageX}, the above can be put in the form appearing in Table \ref{table:analogue}
\begin{equation}\label{eq:Xfluctuation}
     \expvals{\Delta \hat{X}^2}_\beta = \frac{\gamma}{2} \expvals{\hX}_\beta + \expvals{\hX}^2_\beta.
\end{equation}
\par Now that we have the fluctuation of the averaged area density operator, we can discuss the limit to the boundary of the causal diamond. As we mentioned in Section \ref{sec:classicaltheory} and show in Appendix \ref{app:stretchedhorizon}, the area density of the stretched horizon becomes the square of the length of the causal diamond in the limit $h_0 \to 0$. At the level of quantum operators, we can therefore write the area operator of the causal diamond's boundary $\hat{A}$ as
\begin{equation}
    \hat{A} = 4\pi \hat{X} + \mathcal{O}(h_0).
\end{equation}
As we argued in Section \ref{sec:quantumtheory}, the expectation values and fluctuations of the $\hX$ operator do not depend on the relative position of the stretched horizon to the null boundary and are thus independent of $h_0$. Plugging the above in equation \eqref{eq:Xfluctuation} and taking the limit, we obtain
\begin{equation}
    \expvals{\Delta \hat{A}^2}_\beta = 2\pi \gamma \expvals{\hat{A}}_\beta + \expvals{\hat{A}}^2_\beta.
\end{equation}
\par
The above equation constitutes the core result of this paper. It is the direct analogue of Einstein's fluctuation formula for black-body radiation. The first term is the Verlinde--Zurek contribution: it is linear in the mean area and therefore has the shot-noise, or Brownian, scaling characteristic of independent discrete fluctuations. The second term is instead the continuum, or wave-like, contribution, quadratic in the mean area.

The parameter $\gamma$ controls the intrinsic quantum spread of the affine intelligent states. It therefore plays a role analogueous to that of Planck's constant $\hbar$ in the canonical coherent-state description of black-body radiation. Since $\gamma$ carries dimensions of area, it naturally defines the elementary scale of geometric fluctuations. Importantly, $\gamma$ is not fixed by the affine algebra itself. Rather, it labels inequivalent families of intelligent states and should ultimately be determined by the underlying microscopic theory of quantum geometry. In a purely gravitational setting, it is natural to expect $\gamma$ to be of order the Planck area $\ell_P^2$. More generally, however, $\gamma$ should be interpreted as the characteristic quantum-gravity area scale, which could be set by another microscopic length scale, such as $\ell_s^2$ in a string-theoretic realization, or by an effective area gap in a different microscopic completion.

This microscopic area scale controls the limiting regimes of the fluctuation formula. When the average classical area is small compared to $\gamma$, the fluctuations are dominated by the Verlinde--Zurek term
\begin{equation}
    \expvals{\Delta \hat{A}^2}_\beta 
    \simeq 
    2\pi \gamma \expvals{\hat{A}}_\beta,\qquad \bar a = \frac{2G}{\beta\kappa} \ll \frac{\gamma}{2}
\end{equation}
The area fluctuations are then governed by the Brownian-like statistics of microscopic quanta of geometry.
In the opposite large-area, continuum regime
the quadratic contribution dominates and one obtains
\begin{equation}
    \expvals{\Delta \hat{A}^2}_\beta 
    \simeq 
    \expvals{\hat{A}}_\beta^2, \qquad \bar a = \frac{2G}{\beta\kappa} \gg \frac{\gamma}{2}.
\end{equation}
In this regime, the fluctuations are controlled by the collective, continuum behaviour of the gravitational field rather than by the discrete microscopic structure of geometry.
\par
{As in the black-body case, the law of total variance gives a useful interpretation of the two contributions in the thermal fluctuation formula. In analogue with \eqref{eq:totalvariance}, the squared fluctuation of the area operator can be decomposed as
\begin{equation}\label{eq:totalvariancearea}
    \expvals{\Delta \hat{A}^2}_\beta
    =
    \int_{0}^{\infty} \dd a\, P_\beta(a)\,
    \expvals{\Delta \hat{A}^2}_{\gamma,a,b}
    +
    \int_{0}^{\infty} \dd a\, P_\beta(a)\,
    \qty(4\pi a-\expvals{\hat{A}}_\beta)^2 .
\end{equation}
The first term is the thermal average of the intrinsic quantum variance within each intelligent state. Accordingly, it generates the Verlinde--Zurek contribution linear in the mean area. The second term is the classical variance of the expectation values across the thermal ensemble and encodes the thermal spread of the area labels. Accordingly, it generates the wave-like contribution proportional to $\expvals{\hat A}_\beta^2$.}

\par
Finally, let us comment on the relation with the embadon picture introduced in
\cite{Ciambelli:2024swv}. In the null limit $h_0\to 0$, the Minkowski
stretched-horizon symplectic structure reduces to the corresponding spin-0
sector of the null symplectic structure. Under this limit, the stretched-horizon
breathing mode becomes the area density of cuts of the null hypersurface, namely
the field denoted by $\Omega$ in \cite{Ciambelli:2024swv}. This is the same
geometric variable that enters the null pair $(\Omega,\mu)$ and, after
quantization, plays the role of the area-carrying field in the associated
chiral $\beta\gamma$ CFT.
In the continuum null theory, each null generator carries an independent copy of
these degrees of freedom. As a result, the total central charge is formally
proportional to the number of points on the cut and is therefore divergent. The
embadon proposal resolves this by replacing the smooth area density by a
molecular distribution
\begin{equation}
    \hat\Omega(v,\vartheta)
    =
    \sum_{i=1}^{N}
    \hat\Omega_i(v)\,
    \delta^{(d)}(\vartheta-\vartheta_i),
\end{equation}
so that only a finite number of area-carrying null generators remain. The
localized excitations $\hat\Omega_i$ are the embadons, and the total central
charge becomes finite, $c_{\mathrm{tot}}=cN$.
\par
The present result gives concrete support to this picture from an independent
perspective. In \cite{Ciambelli:2024swv}, the molecular replacement of the
continuum area density was introduced as a quantum-gravitational postulate,
motivated by the need to render the null central charge finite. Here, by
contrast, the stretched-horizon analysis shows that the same area density is
selected by the gravitational symplectic structure as a genuine fluctuating
quantum degree of freedom. Moreover, its thermal variance contains a term
linear in the mean area, precisely of the form expected from independent
microscopic constituents. Thus, the present work provides concrete statistical evidence for the embadon picture of \cite{Ciambelli:2024swv}.

\section{Conclusion}\label{sec:conclusion}
In this work we constructed a quantum-statistical analogue of Einstein's
fluctuation argument for black-body radiation in the context of causal-diamond
geometry. Starting from the gravitational phase space associated with a
stretched horizon, we defined a classical theory on the coadjoint orbits of the affine group where the fundamental fields are given by the averaged area density and the average of its radial derivative. Quantizing this system through
affine representations naturally leads to intelligent states in which the fluctuation of the averaged area density operator
scales linearly with the expectation value of the same operator.

Using these states, we constructed a thermal density operator by maximizing the
Shannon entropy under a fixed average horizon energy constraint. In the null limit where the averaged area density becomes the area of the causal diamond's boundary, the resulting
thermal fluctuations satisfy
\begin{equation}\label{eq:mainresult}
    \expvals{\Delta \hat{A}^2}_{\beta}
    =
    2\pi \gamma \expvals{\hat{A}}_{\beta}
    +
    \expvals{\hat{A}}_{\beta}^{2},
\end{equation}
which is the precise analogue of Einstein's fluctuation formula for black-body
radiation.
The structure of this formula is highly suggestive. The quadratic contribution
corresponds to continuum or wave-like fluctuations, while the linear
contribution has the characteristic scaling of independent discrete
excitations. In Einstein's original analysis, the appearance of such a term
provided evidence for the existence of photons prior to the full development of
quantum electrodynamics. In the present case, the analogueous term strongly
suggests the existence of elementary quanta of geometry associated with horizon
area fluctuations.

An important conceptual point is that this conclusion does not rely on the discreteness of
the spectrum of the averaged area density operator. Indeed, the affine operator $\hat X$
has a continuous spectrum. The appearance of particle-like fluctuations instead follows
from the statistical properties of the intelligent states and from the structure of the thermal
distribution in the classical area variable. {This is made explicit by the total variance decomposition of the fluctuations \eqref{eq:totalvariancearea}}. Moreover, the derivation does not require
knowledge of the Hamiltonian of the underlying quantum gravity theory, nor of its spectrum.
In this sense, the present derivation should be understood as a bottom-up argument for
quantum geometry, analogueous to Einstein's original statistical argument for light quanta.
There is therefore no contradiction between the continuous spectrum of $\hat X$ and the
appearance of a fluctuation term characteristic of discrete geometric excitations. The
present construction is not a complete microscopic theory of quantum geometry. It is a
reduced, averaged description in which only the finite-dimensional affine subalgebra of the
stretched-horizon phase space is quantized. In particular, we have averaged over the
stretched-horizon time direction, restricted attention to the spin-0 area mode, and neglected
the quantum fluctuations in the angular directions which would deform the background away
from exact spherical symmetry and, more generally, away from the fixed Minkowski causal
diamond. The continuous spectrum of the averaged affine area operator should therefore
not be interpreted as the final microscopic spectrum of area in the full theory. Rather, the
linear term in the thermal variance should be interpreted as a statistical signature of
underlying geometric quanta, visible already at the level of the reduced phase space.
This is closely analogueous to Einstein's fluctuation argument for black-body radiation. In
that case, the thermodynamic derivation is formulated in terms of a continuous spectral
energy density and does not require prior knowledge of the photon Fock space. Nevertheless,
the appearance of a term linear in the mean energy reveals the particle-like structure of
radiation. In the same spirit, our derivation does not require a discrete area spectrum as an
input. It shows instead that the affine quantum-statistical structure of the horizon area
already contains a Brownian, shot-noise contribution of the type expected from independent
microscopic constituents. The original statistical derivation of Einstein's fluctuation formula
is recalled in Appendix~\ref{app:statisticalderivation}, and the present construction should be
read as its gravitational analogue.

The central mechanism behind the core result of this paper \eqref{eq:mainresult} is the affine structure of the gravitational phase
space. This structure is not merely a technical tool for quantization; it is physically
responsible for the form of the fluctuation law. Since area is a positive quantity, the natural
conjugate transformation is not an additive translation of the area, but a dilation. This is
encoded in the affine commutation relation
\begin{equation}
    [\hat D,\hat X] = -i \hat X ,
\end{equation}
whose uncertainty relation is intrinsically multiplicative
\begin{equation}
  \expvals{\Delta \hat X^2} \expvals{\Delta \hat D^2}
    \geq
    \frac{1}{4}\expvals{\hat X}^2 .
\end{equation}
Thus the scale of the quantum uncertainty is itself controlled by the expectation value of
the area. The linear Verlinde--Zurek term in the thermal variance is therefore tightly
connected to the positivity as was already noted in \cite{Ciambelli:2025flo}. This places the result in the broader conceptual family of affine quantization \cite{Klauder:1999ngq,Watson:2000gac,Klauder:2011uva} while giving these structures a direct geometrical
origin in the stretched-horizon phase space.

The analysis is remarkably general. No specific microscopic theory of quantum gravity
was assumed. The derivation only relies on the affine structure of the stretched-horizon
phase space, the existence of intelligent states saturating the affine uncertainty relation, and
standard principles of statistical mechanics. This universality suggests that the fluctuation
formula may capture generic properties of quantum gravitational horizons independently
of the detailed ultraviolet completion. This is reminiscent of the recent work on the \textit{corner proposal} which provides a bottom-up approach to quantum gravity \cite{Freidel:2015gpa,Freidel:2020xyx,Freidel:2020svx,Ciambelli:2021vnn,Freidel:2021cjp,Ciambelli:2021nmv,Ciambelli:2022cfr,Ciambelli:2022vot,Freidel:2023bnj,Ciambelli:2023bmn,Ciambelli:2024swv,Ciambelli:2024qgi,Varrin:2024sxe,Neri:2025fsh,Varrin:2025okc,Kowalski-Glikman:2025hom,Ciambelli:2025ztm,Varrin:2026bmb}.

The parameter $\gamma$ controls the intrinsic quantum spread of the area fluctuations and
naturally defines the characteristic scale of geometric quanta. A compelling possibility is
that $\gamma$ is ultimately fixed by the microscopic theory and is of order the Planck area.
More generally, however, $\gamma$ should be interpreted as the characteristic area scale of
the ultraviolet completion. Depending on the microscopic theory, this scale could coincide
with $\ell_P^2$, with a string scale $\ell_s^2$, or with an effective area gap associated with
the elementary quantum-geometric degrees of freedom.

Our results also provide strong support for the embadon picture proposed in \cite{Ciambelli:2024swv}.
In that framework, null geometry admits elementary excitations carrying quanta of area.
The present work shows that thermal area fluctuations exhibit precisely the statistical
structure expected for an ensemble of such geometric quanta.

This perspective is also directly relevant to the Verlinde--Zurek program on observable
spacetime fluctuations. A central feature of that program is that quantum fluctuations
controlled by microscopic area scales can induce correlations at macroscopic distances,
providing a form of UV/IR mixing potentially accessible to interferometric experiments.
Within the framework developed here, the observation of a fluctuation signal with the
Verlinde--Zurek scaling would have a particularly sharp interpretation. It would not merely
indicate the presence of an additional stochastic noise source, but would provide evidence
for the quantum-statistical structure of spacetime itself: namely, for area fluctuations whose
linear variance term is tied to the affine quantum geometry of horizons. In this sense,
interferometric evidence for Verlinde--Zurek-type fluctuations would constitute strong
experimental support for the quantum nature of spacetime.
\par
We close by mentioning several future directions opened by the present work. A natural next step is to investigate whether the full Planck law for area fluctuations can be recovered by determining the density of states associated with a single area mode, and to clarify possible connections with gravitational-wave physics. It would also be interesting to explore the relation of our framework to other thermal descriptions of causal diamonds and related approaches developed in the literature (see e.g., \cite{Jacobson:2022gmo}, \cite{Arzano:2026vwz} and references therein). Finally, the present analysis has been restricted to the Minkowski background, and its extension to more general spacetimes constitutes an important direction for future research.

\section*{Acknowledgment} 

We dedicate this paper to the memory of Rob Leigh (1964–2026).

\noindent
We thank Luca Cambelli, Laurent Freidel, Djordje Minic, and Giulio Neri for many discussions.

\appendix
\section{Statistical Derivation of Energy Fluctuation}\label{app:statisticalderivation}
The energy density of the radiation of a black-body depends on the frequency of light emitted through Planck's law (in units where $c=k_B = 1$)
\begin{equation}\label{eq:plancklaw}
    \rho(\nu) = \frac{8\pi h \nu^3}{e^{\frac{h\nu}{T}}-1},
\end{equation}
where $h$ is Planck's constant and $T$ is the temperature of the black-body. For low frequencies this reduces to
\begin{equation}
    \rho(\nu) \approx 8\pi \nu^2 T,\qquad \nu \ll 1
\end{equation}
which is the Rayleigh--Jeans law. For high frequencies $\nu\gg 1$, the exponential dominates and we obtain Wien's law instead
\begin{equation}
    \rho(\nu) \approx 8\pi h \nu^3 e^{-\frac{h \nu}{T}}, \qquad \nu \gg 1.
\end{equation}
The mean energy of the radiation in a narrow frequency range $(\nu,\nu + \Delta \nu)$ in a small volume $V$ is given by 
\begin{equation}\label{eq:meanenergy}
    \bar{E}(\nu,T,V) = \rho(\nu,T) V\Delta \nu.
\end{equation}
In order to compute the mean fluctuations, we can use the following formula
\begin{equation}\label{eq:meanfluctuations}
    \overline{\Delta E^2} = T^2 \dv{\bar{E}}{T}.
\end{equation}
While the canonical ensemble gives a straightforward derivation of the above equation, it is not adapted to the description of black body radiations. In his 1909 paper \cite{einstein1909gegenwartigen}, Einstein rederived this equation using the Boltzmann equation for the entropy instead. We present the argument here, closely following \cite{DuncanJanssen}.
Consider a box of volume $V = V_0 + \delta V$ with perfectly reflecting walls filled with black-body radiation. In the small subvolume $\delta V$, define the instantaneous energy fluctuation
\begin{equation}
    \epsilon = E -\overline{E}.
\end{equation}
 Since the walls are perfectly reflecting, the fluctuation of the energy in the volume $V_0$ is given by
 \begin{equation}
     -\epsilon = E_0 -\overline E_0.
 \end{equation}
The average squared fluctuation of the energy in $\delta V$ is defined by
\begin{equation}
    \overline{\Delta E^2} = \frac{\int_{-\infty}^\infty \epsilon^2 W(\epsilon)\dd\epsilon}{\int_{-\infty}^\infty W(\epsilon)\dd\epsilon},
\end{equation}
 where the probability of the instantaneous fluctuation $\epsilon$ can be expressed through Boltzmann's equation
 \begin{equation}
     W(\epsilon)\dd\epsilon = e^{\Delta S(\epsilon)}\dd\epsilon.
 \end{equation}
 The variation of the entropy can be computed at next-to-leading order
 \begin{equation}\label{eq:entropyexpansion}
     \Delta S(\epsilon) = \qty(\overline{\pdv{S}{E}} - \overline{\pdv{S_0}{E_0}})\epsilon + \frac12 \qty(\overline{\pdv[2]{S}{E}}+ \overline{\pdv[2]{S_0}{E_0}})\epsilon^2 + \mathcal{O}(\epsilon^3).
 \end{equation}
Using the standard thermodynamical relation we can write
\begin{equation}
    \overline{\pdv[2]{S}{E}} = \overline{\frac{\partial}{\partial E}\qty(\frac{1}{T})} = \frac{\partial}{\partial T} \qty(\frac{1}{T})\overline{\pdv{T}{E}} = -\frac{1}{T^2 C_{\delta V}}.
\end{equation}
 where we have defined the heat capacity at fixed volume
 \begin{equation}
     C_V = \dv{\overline E}{T}.
 \end{equation}
Since the subsystem is in thermal equilibrium with the rest of the system, we also have
\begin{equation}
     \overline{\pdv[2]{S}{E}} = -\frac{1}{T^2 C_{V_0}}.
\end{equation}
 Plugging these expressions back into equation \eqref{eq:entropyexpansion} and using the fact that $C_{V_0}\gg C_{\delta V}$ we obtain
 \begin{equation}
     \Delta S(\epsilon) = - \frac{\epsilon^2}{2 T^2 C_{\delta V}} + \mathcal{O}(\epsilon^3).
 \end{equation}
Finally, we insert this result back into the average squared fluctuations of the energy to obtain
\begin{equation}
    \overline{\Delta E^2} = T^2 \dv{\overline{E}}{T},
\end{equation}
which is the formula stated above. \par
 Using equations \eqref{eq:meanenergy} and \eqref{eq:meanfluctuations}, we can write
\begin{equation}
    \overline{\Delta E^2} = T^2 \pdv{\rho}{T} V \Delta \nu.
\end{equation}
Using the explicit low frequency Rayleigh--Jeans expression for $\rho(\nu,T)$, we find
\begin{equation}
    \overline{\Delta E^2} = T^2 8 \pi \nu^2 V \delta \nu = \frac{1}{8\pi \nu^2}\frac{\overline{E}^2}{V\delta\nu},\quad h\nu <<T
\end{equation}
i.e. the mean fluctuation of the energy scales like the mean energy. In the high-frequency regime dictated by Wien's law we find instead 
\begin{equation}
    \overline{\Delta E^2} \approx  h \nu \overline{E}, \quad h \nu >>T
\end{equation}
which is the "signature of particle" behaviour.
\section{Stretched Horizon}\label{app:stretchedhorizon}
We start with the standard Minkowski line element in spherical coordinates,
\begin{equation}
    \dd s^2=-\dd t^2+\dd r^2+r^2\dd\Omega_2^2 .
\end{equation}
The causal diamond of size $L$ is defined as the region $r-L \leq t\leq L-r$. It is convenient to parametrize it in the retarded and advanced null coordinates
\begin{equation}
    u=t-r,
    \qquad
    v=t+r,
\end{equation}
for which the metric becomes
\begin{equation}
    \dd s^2
    =
    -\dd u\,\dd v
    +
    \qty(\frac{v-u}{2})^2\dd\Omega_2^2 .
\end{equation}
The causal diamond is then described by the region
$ -L\leq u\leq v\leq L$ and its past and future null boundaries are easily described in the $u,v,\theta^A$ coordinates
\begin{equation}
    H_-= (-L,v,\theta^A),
    \qquad
    H_+=(u,L,\theta^A).
\end{equation}
Their intersection $(u=-L,v=L)$ is the bifurcation two-sphere
which corresponds to $t=0$ and $r=L$. One can define Gaussian-null coordinates $(\tau,\tilde{r})$ adapted to the future horizon to describe the interior of the causal diamond \cite{Bub:2024qmc} 
\begin{equation}
    u(\tau)
    =
    -L+\frac{1}{\kappa}e^{\kappa\tau+\alpha},
    \qquad
    v(\tau,\tilde r)
    =
    L-2\tilde r\,e^{-\kappa\tau-\alpha},
\end{equation}
where $\kappa$ is a spacetime constant whose meaning will soon become clear. The coordinate $\tilde{r}$ is a transverse coordinate measuring the distance to the future horizon and the constant $\tau$ slices are outgoing null hypersurfaces of constant $u$. The bifurcating sphere is located at $\tilde{r} = 0$ and $\tau \to -\infty$. In these coordinates,
the line element becomes
\begin{Align}
    \dd s^2 &= -2\kappa \tilde{r}\dd\tau^2 + 2  \dd \tau \dd \tilde{r} + \varphi(\tau,\tilde{r}) \dd\Omega_2^2,\\
    \varphi(\tau,\tilde{r}) &= \qty(L - \frac{1}{2\kappa}e^{\kappa\tau + \alpha}- \tilde{r} e^{-\kappa \tau - \alpha})^2.
\end{Align}
\par
While Gaussian-null coordinates adapted to the past horizon $H_-$ can also be constructed, we would like a coherent description of the entire boundary of the causal diamond. This can be achieved through the stretched horizon construction. To that end, we define the new coordinate
\begin{equation}
    \tilde{r} = \rho + \frac{h_0}{\kappa},
\end{equation}
such that the stretched horizon $H_s$ is located at $\rho = 0$ and $h_0$ is a dimensionless parameter that defines its distance to the null horizon.
The Minkowski line element then becomes
\begin{Align}\label{eq:stretchedhorizonmetric}
    \dd s^2 &=  -2(h_0 + \kappa \rho) \dd \tau^2 + 2 \dd \tau \dd \rho + \varphi(\tau,\rho)\dd\Omega_2^2,\\
  \varphi(\tau,\rho)
&=
\left(
L-\frac{1}{2\kappa}e^{\kappa\tau+\alpha}
-\left(\rho+\frac{h_0}{\kappa}\right)e^{-\kappa\tau-\alpha}
\right)^2 .
\end{Align}
The induced metric on the stretched horizon is given by
\begin{equation}
    \dd s^2\eval_{H_s} = -2 h_0 \dd \tau^2 + \varphi(\tau,0)\dd\Omega_2^2.
\end{equation}
Time evolution on the stretched horizon is generated by the vector $l = \partial_\tau$ with norm
\begin{equation}
    l^2 = -2 h_0.
\end{equation}
The constant $\kappa$ is the inaffinity of the clock vector on the stretched horizon. This can be seen as follows. The relevant Christoffel symbols of the metric \eqref{eq:stretchedhorizonmetric} are
\begin{equation}
    \Gamma_{\tau\tau}^\tau = \kappa, \qquad \Gamma_{\tau\tau}^\rho = 2 \kappa (h_0 + \kappa \rho).
\end{equation}
We can then compute
\begin{equation}
    l^\mu \nabla_\mu l = \kappa \partial_{\tau} + 2 \kappa h_0 \partial_\rho.  
\end{equation}
Applying the projector to the stretched-horizon $\pi_{H_s}(\partial_\tau) = \partial_\tau$ and $\pi_{H_s}(\partial_\rho)= 0$, we obtain
\begin{equation}
    \pi_{H_s}\qty(l^\mu \nabla_\mu l^\nu) = \kappa l^\nu,
\end{equation}
which is the defining equation of inaffinity. We can compute the derivative of the clock vector norm squared
\begin{equation}
    \partial_\mu\qty(l^2) = -2 \kappa l_\mu - 4 \kappa h_0 \partial_\mu \tau.
\end{equation}
In the null-horizon limit, we therefore have
\begin{equation}
    \lim_{h_0 \to 0}\partial_\mu\qty(l^2) = -2 \kappa l_\mu,
\end{equation}
which is the definition of the surface gravity on the causal diamond's null boundary. Note that this definition depends on the normalization of the clock vector.
\par
We now compute the time average of $\varphi(\tau)$ and show that it corresponds to $\frac{A}{4\pi}$ in the $h_0\to 0$ limit. We can compute the average of the area density explicitly
\begin{equation}\label{eq:integralexpansion}
    \bar{\varphi} =  L^2 + \frac{h_0}{\kappa^2} + \frac{\sinh(\kappa T)}{T\kappa^3}\qty(\frac14e^{2\bar{u}} + h_0^2 e^{-2\bar{u}}) - \frac{2L}{\kappa^2}\frac{\sinh(\frac{\kappa T}{2})}{T}(e^{\bar{u}} + 2 h_0 e^{-\bar{u}}),
\end{equation}
where we defined $\bar{u} = \alpha + \frac{\kappa}{2}(\tau_+ + \tau_-)$. In order to go further, we must compute the time $T$. The endpoints of the stretched horizon are defined by the presence of caustics
\begin{equation}\label{eq:endpoints}
    \varphi(\tau_\pm)=
\left(
L
-
\frac{1}{2\kappa}e^{\kappa \tau_\pm+\alpha}
-
\frac{h_0}{\kappa}e^{-\kappa \tau_\pm-\alpha}
\right)^2  = 0.
\end{equation}
Denoting
\begin{equation}
    y_\pm = e^{\kappa\tau_\pm + \alpha},
\end{equation}
the equation becomes
\begin{equation}
    y^2_\pm - 2 \kappa L y_\pm + 2 h_0 = 0,
\end{equation}
from which one can deduce
\begin{equation}
    y_-y_+   =2 h_0 = e^{2\bar{u}} \Rightarrow \bar{u} = \frac12 \ln(2h_0).
\end{equation}
Equation \eqref{eq:integralexpansion} then becomes
\begin{equation}
    \bar\varphi = L^2 + \frac{h_0}{\kappa^2} + \frac{\sinh(\kappa T)}{T\kappa^3}h_0 - \frac{4L}{\kappa^2}\frac{\sinh(\frac{\kappa T}{2})}{T}(\sqrt{2h_0}).
\end{equation}
Solving the equation for the endpoints gives
\begin{align}
    \sinh(\kappa T) &= \frac{\kappa L}{h_0}\sqrt{\kappa^2 L^2 -2 h_0},\\
    \sinh(\frac{\kappa T}{2}) &= \frac{1}{\sqrt{2h_0}} \sqrt{\kappa^2 L^2 - 2 h_0}.
\end{align}
Plugging this result back into the expression for the average area density yields
\begin{equation}
    \bar\varphi = L^2 + \frac{h_0}{\kappa^2}-\frac{3 L}{2 \kappa }\frac{\sqrt{\kappa^2 L^2 -2 h_0}}{\mathrm{arcosh}\qty(\frac{\kappa L}{\sqrt{2h_0}})}.
\end{equation}
From the above it is easy to see that
\begin{equation}
    \lim_{h_0 \to 0}\bar \varphi = L^2 = \frac{A}{4\pi}.
\end{equation}

\section{Affine Coadjoint Orbits}\label{app:coadjointorbits}
The classical phase space associated with the averaged fields \eqref{eq:spin0fields}
can be described on the coadjoint orbits of the affine group that can be described as follows. Consider the matrix representation of $\mathrm{Aff}_+(\RR)$ seen as a subgroup of $\mathrm{GL}\qty(2,\RR)$
\begin{equation}\label{eq:groupelement}
    (q,p) \rightarrow \mqty(q^{-1}&-p\\0&1).
\end{equation}
The corresponding Lie algebra can be written
\begin{equation}
    \mathfrak{aff}(\RR) = \qty{\mqty(-a&b&\\0&0)\mid a,b \in \RR},
\end{equation}
and is generated by
\begin{equation}
    X = \mqty(0&1\\0&0),\qquad D = \mqty(-1&0\\0&0).
\end{equation}
Any group element \eqref{eq:groupelement} can then be written
\begin{equation}
    g(q,p) = \exp(-p X)\exp(\ln(q)D).
\end{equation}
The coalgebra can be written
\begin{equation}
    \mathfrak{aff}^*(\RR) = \qty{\mqty(\alpha&0\\\beta&0)\mid \alpha,\beta\in\RR}.
\end{equation}
Using the standard pairing $\braket{A^*}{A} = \Tr[A^*A]$, we can compute the dual basis
\begin{equation}
    X^* = \mqty(0&0\\1&0),\qquad D^* =\mqty(-1&0\\0&0).
\end{equation}
We can compute the adjoint action
\begin{equation}
    \mathrm{Ad}_{g(q,p)}X = g(q,p) X g^{-1}(q,p) =  q^{-1}X,\quad \mathrm{Ad}_{g(q,p)}D = D -pX,
\end{equation}
From which we deduce the coadjoint action
\begin{equation}
    \mathrm{Ad}^*_{g(q,p)}X^* = q X^* + qp D^*,\quad \mathrm{Ad}^*_{g(q,p)}D^* = D^*.
\end{equation}
This implies that a general point $\mu = x X^* + d D^* \in \mathfrak{aff}(\RR)^*$ transforms under the coadjoint action as 
\begin{equation}\label{eq:coadjointaction}
    \mu(x,d) \mapsto \mu(x',d'), \quad x' = q x, \, d' = d + qpx.
\end{equation}
It is easy to see that the coalgebra $\mathfrak{aff}(\RR)^* \cong \RR^2$ foliates into three classes of orbits
\begin{itemize}
    \item The upper half plane $\mathcal{O}_{\mathrm{Aff}}^+ = \qty{(x,d)\in \RR^2 \mid x>0}$
    \item The lower half plane $\mathcal{O}_{\mathrm{Aff}}^- = \qty{(x,d)\in \RR^2 \mid x<0}$
    \item The singleton orbits $\mathcal{O}_{\mathrm{Aff}}^0 = \qty{(0,d) \in \RR^2 \mid d\in \RR}$
\end{itemize}
Since the average area density is associated with the coordinate $x$, the coadjoint orbit of interest is the upper half plane $\mathcal{O}_{\mathrm{Aff}}^+$. The fundamental vector fields can be computed in the $(x,d)$ coordinates
\begin{equation}
    \xi_X = -x \partial_d, \qquad \xi_D = x\partial_x.
\end{equation}
The KKS form at a point $\mu = x X^* + d D^* \in \mathcal{O}_{\mathrm{Aff}}^+$ is given by 
\begin{equation}
    \Omega_{\mathrm{KKS}}^\mu(\xi_X,\xi_D) = \braket{\mu}{[X,D]} = x.
\end{equation}
Let us write $\Omega_{\mathrm{KKS}} = f(x,d)\dd x \wedge \dd d$. Using the fundamental vector fields, we compute 
\begin{equation}
    \Omega_{\mathrm{KKS}}(\xi_X,\xi_D) = x^2 f(x,d).
\end{equation}
Comparing with the result above, we conclude
\begin{equation}
    \Omega_{\mathrm{KKS}} = \frac{1}{x}\dd x \wedge \dd d.
\end{equation}
The main text also introduces Darboux $Q,P$ defined as $x=Q,d = QP$, for which the symplectic form takes its canonical form
\begin{equation}
    \Omega_{\mathrm{KKS}} = \dd Q \wedge \dd P.
\end{equation} 
This is the familiar result that the half-plane cotangent bundle $T^* \RR_+$ has an affine symmetry generated by the functions
\begin{equation}
    F_x(Q,P) = Q, \qquad F_d(Q,P) = \frac12(PQ+QP),
\end{equation}
The use of capital letters is to differentiate the coadjoint orbit coordinates from the group manifold coordinates, although they are of course isomorphic.
The coadjoint orbit simply has the additional structure of a symplectic bracket. In fact, the coadjoint action reproduces the group law
\begin{equation}
   \mathrm{Ad}^*_{g(q,p)} Q = qQ,\quad \mathrm{Ad}^*_{g(q,p)} P = \frac{P}{q}+p. 
\end{equation}
Note that this is nothing else than the standard isomorphism
\begin{equation}\label{eq:isomorphism}
    \mathcal{O}^{\mu}_{\mathrm{Aff}} = \mathrm{Aff}_+(\RR)\slash G_m,
\end{equation}
where $G_m$ is the stabilizer subgroup of the point $m\in\mathfrak{aff}^*_+(\RR)$. It is clear from the action \eqref{eq:coadjointaction} that the stabilizer is trivial for any $m\in\mathcal{O}_{\mathrm{Aff}}^+$. Equation
\eqref{eq:isomorphism} then becomes an isomorphism between the coadjoint orbit and the group given by the map
\begin{Align}
    \phi: \mathcal{O}_{\mathrm{Aff}}^+ &\longrightarrow \mathrm{Aff}_+(\RR)\\
    (Q,P)&\longmapsto \phi(Q,P) = (q=Q,p=P).
\end{Align}

\bibliographystyle{JHEP}
\bibliography{refs}

\end{document}